\documentclass[showpacs,prd,nofootinbib,reprint,preprintnumbers]{revtex4-1}
\usepackage{adjustbox}
\usepackage[utf8]{inputenc}
\usepackage{color,graphicx,epsfig}
\usepackage{amsmath,amssymb,dsfont,epsfig,graphicx,xcolor,fontenc}
\usepackage{ifpdf}
\usepackage{amsmath}
\usepackage{bm}
\usepackage[english]{babel}
\usepackage{graphicx}
\usepackage{cancel}
\usepackage{amsfonts}
\usepackage{amssymb}
\usepackage{braket}
\usepackage{hyperref}
\usepackage{epstopdf}
\usepackage{ulem}
\definecolor{nicered}{rgb}{0.7,0.1,0.1}
\definecolor{nicegreen}{rgb}{0.1,0.5,0.1}
\definecolor{violet}{rgb}{0.7,0.3,0.3}
\hypersetup{colorlinks,citecolor= nicegreen,linkcolor= nicered}

\newcommand{\re}{\mathrm{Re}}
\newcommand{\im}{\mathrm{Im}}
\newcommand{\mc}{\mathcal}
\newcommand{\mrm}{\mathrm}
\newcommand{\e}[1]{\,\mathrm{#1}}
\newcommand{\yuu}{Y_{u\bar{u}}}
\newcommand{\ycc}{Y_{c\bar{c}}}
\newcommand{\yqq}{Y_{q\bar{q}}}

\def\LjubljanaFMF{Faculty of Mathematics and Physics, University of Ljubljana,
 Jadranska 19, 1000 Ljubljana, Slovenia }
\def\LjubljanaIJS{Jo\v zef Stefan Institute, Jamova 39, 1000 Ljubljana, Slovenia}
\def\IFIC{Instituto de F\'isica Corpuscular,
Universitat de Val\'encia – Consejo Superior de Investigaciones Cient\'ificas,
Parc Cient\'ific, E-46980 Paterna, Valencia, Spain}
\def\Bari{Istituto Nazionale di Fisica Nucleare, Sezione di Bari, Via Orabona 4, 70126 Bari, Italy}

\newcommand{\Cc}{\ensuremath{\mathcal{C}}}

\begin{document}
\preprint{BARI-TH/756-24}

\title{CP-odd window into long distance dynamics in rare semileptonic $B$ decays}

\author{Jernej~F.~Kamenik}
\email[Electronic address:]{jernej.kamenik@cern.ch} 
\affiliation{\LjubljanaIJS}
\affiliation{\LjubljanaFMF}

\author{Nejc Ko\v snik}
\email[Electronic address:]{nejc.kosnik@ijs.si} 
\affiliation{\LjubljanaIJS}
\affiliation{\LjubljanaFMF}

\author{Mart\'in Novoa-Brunet}
\email[Electronic address:]{martin.novoa@ific.uv.es}
\affiliation{\IFIC}
\affiliation{\Bari}
\begin{abstract}
We consider the combined measurements of CP-averaged decay rates and direct CP asymmetries of $B^\pm\to K^\pm \ell^+ \ell^-$ and $B^\pm\to \pi^\pm \ell^+ \ell^-$ to probe (non-local) four-quark operator matrix element contributions to rare semileptonic B meson decays. We also explore how their effects could be in principle disentangled from possible local new physics effects using $U$-spin relations. To this end, we construct a ratio of CP-odd decay rate differences which are exactly predicted within the standard model in the $U$-spin limit, while the leading $U$-spin breaking effects can also be systematically calculated. Our results motivate binned measurements of the direct CP asymmetry in $B^\pm\to \pi^\pm \ell^+ \ell^-$ as well as dedicated theoretical estimates of $U$-spin breaking both in local form factors as well as in four-quark matrix elements.
\end{abstract}

\maketitle

%
\section{Introduction}
%

Over the past decade, the LHCb experiment has produced several intriguing results on rare semileptonic decays of $b$-flavoured hadrons~\cite{LHCb:2014cxe, LHCb:2015svh, LHCb:2015tgy, LHCb:2015wdu, LHCb:2016due, LHCb:2019efc, LHCb:2020lmf, LHCb:2020gog, LHCb:2021zwz, LHCb:2021lvy, LHCb:2022vje, LHCb:2022qnv}. In particular, the measurements of several decay rates as well as angular observables in $B\to K^{(*)} \mu^+ \mu^-$ exhibit persistent tensions with current theoretical estimates with the standard model (SM)~\cite{Alguero:2023jeh,Capdevila:2023yhq}.

The theory of rare semileptonic decays unfortunately suffers from substantial hadronic uncertainties. The dominant contributions to the decay amplitudes can be divided into matrix elements of local (quark field bilinear) operators and non-local matrix elements of four-quark operators contracted with the electromagnetic (EM) current~\cite{Beneke:2001at, Beneke:2009az}. There has been tremendous progress in precision evaluations of the former using Lattice QCD methods~\cite{Parrott:2022rgu}. On the other hand, a robust theoretical estimation of the latter is still beyond reach, despite enduring efforts~\cite{Beneke:2000wa, Beneke:2001at, Beneke:2004dp, Khodjamirian:2010vf, Khodjamirian:2012rm, Hambrock:2015wka, Khodjamirian:2017fxg,Gubernari:2020eft,Gubernari:2022hxn,Gubernari:2023puw}. 

In the next decade, large datasets of both LHCb~\cite{LHCb:2018roe} as well as Belle II~\cite{Belle-II:2018jsg} experiments are expected to provide more detailed and precise measurements of rates, spectra as well as angular observables and CP asymmetries in both $b\to s$ and $b \to d$ semileptonic transitions. If the current intriguing results are confirmed and strengthened, it will be imperative to disentangle possible explanations in terms of unaccounted for hadronic effects from possible signals of physics beyond the standard model~(BSM), see e.g. refs.~\cite{Ciuchini:2020gvn, Gubernari:2020eft, Descotes-Genon:2020tnz,Gubernari:2022hxn, Marshall:2023aje, Bordone:2024hui} for some recent proposals.

In the present work we explore how the measurements of (direct) CP asymmetries in $B^\pm\to K^\pm \ell^+ \ell^-$ and $B^\pm\to \pi^\pm \ell^+ \ell^-$ could play a crucial role in this endeavour.

In particular,
direct CP violation arises from the interference of decay amplitudes with different CP-odd as well as CP-even phases.
While local (short distance) SM as well as possible BSM amplitudes in these $B \to K \ell\ell$ and $B \to \pi \ell \ell$ processes can carry different CP-odd phases, there can be no significant CP-even phase differences between them.\footnote{Possibly observable EM rescattering effects in presence of large BSM $b\to s \tau^+\tau^-$ amplitudes have recently been considered in ref.~\cite{Cornella:2020aoq}.} Any signal of direct CP-violation is thus necessarily proportional to the absorptive parts of some long-distance rescattering contributions. In the SM, these are dominated by non-local four-quark operator matrix elements and have precisely predicted CP-odd phases relative to the computable short distance amplitudes.  

In the following we demonstrate how future combined measurements of CP-averaged decay rates and direct CP-asymmetries of $B^\pm\to K^\pm \ell^+ \ell^-$ and $B^\pm\to \pi^\pm \ell^+ \ell^-$ transitions can be used to learn about the sizes of the four-quark operator matrix elements and how their effects could be in principle disentangled from possible local BSM effects. As a byproduct, we construct a ratio of CP-odd decay rate differences of $B^\pm\to K^\pm \ell^+ \ell^-$ and $B^\pm\to \pi^\pm \ell^+ \ell^-$ which is exactly predicted in the $U$-spin limit and within a certain kinematical regime. We estimate this ratio in presence of know $U$-spin breaking effects due to kinematics and differences in local operator matrix elements (form factors).

The rest of the paper is structured as follows: In Sec.~\ref{sec:decomp} we decompose the CP-even and CP-odd $B\to P \ell^+\ell^-$ rates in terms of local and long-distance amplitudes. We apply this decomposition to specific $b\to s$ and $b \to d$ transitions in Sec.~\ref{sec:BKpi} and discuss the CKM and kinematics induced hierarchies of different contributions. Next, in Sec.~\ref{sec:combine} we discuss how to combine information from both flavor modes using $U$-spin and project the resulting sensitivity of possible future LHCb and Belle II measurements to short and long-distance effects. Finally, we summarize our results in Sec.~\ref{sec:conclusions}.  

%
\section{Decomposition of CP structure in $B\to P \ell^+\ell^-$}
\label{sec:decomp}
%

In the SM, the differential decay rate of $B\to P \ell^+\ell^-$, mediated by $b\to q' \ell^+ \ell^-$ quark-level amplitudes, with $\ell = e,\mu$ and $q'=s(d)$ for $P=K(\pi)$, can be written as~\cite{Bobeth:2007dw}
\begin{align}
\begin{split}
    \frac{d\Gamma_P}{dq^2}&=\mathcal{N}_P \,\left(f^{(P)}_+\right)^2\,\left(|\Cc_{10}|^2 
    + \left|\Cc_9^\mrm{eff} + \tilde{f}^{(P)}_T \Cc_7\right|^2
    \right)\,.
    \end{split}
\end{align}
Here $q^2 \equiv (p_{\ell^+}+p_{\ell^-})^2$ and we have neglected terms of $\mc{O}(m_\ell^2)$.\footnote{We consistently set $m_\ell = 0$ in all subsequent expressions.} In the above expression, $f^{(P)}_+(q^2)$ is the form factor for the vector-current $B \to P$ matrix element while $\tilde f^{(P)}_T\equiv 2 f^{(P)}_T(q^2) ({m_b+m_{q'}})/f^{(P)}_+(q^2) ({m_B + m_P})$ is a ratio of tensor-to-vector form factors. We use the conventional definition of the form factors (as in e.g.~\cite{Beneke:2000wa}) and employ lattice QCD results for $B \to \pi$~\cite{Leljak:2021vte} and $B\to K$~\cite{Gubernari:2023puw} transitions.
Correspondingly, $\Cc_i$ are the relevant local operator Wilson coefficients  with short-distance SM values $\Cc^{\rm SM}_{10} = -4.31$, $\Cc^{\rm SM}_7 = -0.292$ and $\Cc^{\rm SM}_9 = 4.07$ at the scale $\mu_b
 = 4.8$ GeV~\cite{Descotes-Genon:2013vna}. We assume that short distance $\Cc_{7,9,10}$ are real throughout the paper unless explicitly stated otherwise. Finally, the $q^2$-dependent normalisation factor $\mc{N}_P$ reads
\begin{align}
    \mc{N}_P & = \frac{G_F^2 \alpha^2 |\lambda_t^{(q')}|^2}{3\cdot 512\pi^5 m_B^3} \,\lambda_P^{3/2}(q^2)\,,
\end{align}
with $\lambda_P(q^2) = (m_B^2 +m_P^2 +q^2)^2 - 2(m_B^2 m_P^2 + m_B^2 q^2 + m_P^2 q^2)$, while $G_F$ and $\alpha$ are the Fermi constant and the fine structure constant, respectively. The normalisation $\mc{N}_P$ also contains a CKM factor $\lambda_{q''}^{(q')}=V_{q''b} V_{q''q'}^*$. 
 
Contributions from four-quark operators can be taken into account by effectively modifying $\Cc_9$ as follows
\begin{equation}
\begin{split}
\Cc_9^{\rm eff}(q^2)&=\Cc_9 - \tilde\lambda_c^{(q')} \ycc(q^2)- \tilde\lambda_u^{(q')} \yuu(q^2)\\
    & \phantom{=}+Y_{d\bar d}(q^2)+Y_{s\bar s}(q^2)\,,
\end{split}
\end{equation}
where $Y_{q\bar q}(q^2)$ parametrize relative effects due to $q\bar q$ rescattering amplitudes. In general they depend on the $q''$ flavor as well as the final state $P$ with flavor $q'$ . We leave these dependencies implicit and return to this point in Sec.~\ref{sec:combine}. In the above expression we have introduced $\tilde\lambda_{q''}^{(q')} = \lambda_{q''}^{(q')}/\lambda_{t}^{(q')}$ and we will repeatedly employ the unitarity of the CKM, $\tilde \lambda_u^{(q')}+\tilde \lambda_c^{(q')}+1=0$, in the following.
We can safely take $Y_{d\bar d}(q^2)\to 0$, $Y_{s\bar s}(q^2)\to 0$ due to tiny Wilson coefficients~\cite{Buchalla:1995vs}. Short distance contributions to $Y_{q\bar q}(q^2)$ can be computed perturbatively~\cite{Buchalla:1995vs}, however, for the remaining $\ycc$ and $\yuu$ we do not factorize the short and long distance~(LD) effects. Then the CP-averaged decay rate and the CP-odd rate difference read
\begin{widetext}    
\begin{align}
    \label{eq:CPAvgRate}
    \frac{d(\Gamma_P+\bar\Gamma_P)/2}{dq^2}
    &= 
    \mc{N}_P \left(f^{(P)}_+\right)^2 \left[\Cc_{10}^2 + (\Cc_9+\tilde{f}^{(P)}_T \Cc_7)^2 -2 (\Cc_9 +\tilde{f}^{(P)}_T \Cc_7)
    \left\{\re (\tilde\lambda_c^{(q')})\,(\re \ycc - \re \yuu) - \re \yuu\right\}  \right.\nonumber\\
    &\left.\phantom{=\mc{N}_P \left(f^{(P)}_+\right)^2} -2\left(|\tilde \lambda_c^{(q')}|^2+\re \tilde \lambda_c^{(q')}\right) \re(\ycc \yuu^*) + |\tilde \lambda_c^{(q')}|^2 |\ycc|^2 + |\tilde \lambda_u^{(q')}|^2 |\yuu|^2 \right]\,,\\
    \frac{d(\Gamma_P-\bar\Gamma_P)}{dq^2} &= 4 \mathcal{N}_P \left(f^{(P)}_+\right)^2 \,\im\tilde\lambda_c^{(q')}\,\left[(\Cc_9+\tilde{f}^{(P)}_T \Cc_7)\left(\im\ycc - \im \yuu\right) + \im  (\ycc \yuu^*) \right] \,. \label{eq:rateDiff-general}
    \end{align}
\end{widetext}
Here we denote by $\Gamma_P$~($\bar \Gamma_P$) the decay rate of $B^{-(+)} \to P^{-(+)} \ell^+ \ell^-$.
We immediately observe that the expression for the CP-odd rate difference is proportional to the absorptive amplitude and thus uniquely probes imaginary parts of $Y_{c\bar c}$ and $Y_{u\bar u}$. Further simplifications to the above expressions can arise due to specific CKM ($\tilde \lambda^{(q')}_{q''}$) hierarchies and in specific kinematical ($q^2$) regions where the absorptive amplitudes are constrained.
We study these effects for specific cases of $B\to K$ and $B\to \pi$ transitions in the next section.

%
\section{$B\to K\ell\ell$ Vs. $B \to \pi \ell\ell$}
\label{sec:BKpi}
%

In the case of $b \to s$ transition the overall CKM factor using the Wolfenstein expansion up to $\mathcal O(\lambda^4)$ is $\lambda_t^{(s)} = -A \lambda^2 + A \lambda^4 (1/2 +i \eta-\rho)$, whereas $\ycc$ and $\yuu$ enter the amplitude with relative CKM factors
\begin{align}
    \tilde \lambda_c^{(s)} &= -1 + \lambda^2 (\rho - i \eta) + \mc{O}(\lambda^4) \,,\\
    \tilde \lambda_u^{(s)} &= -\lambda^2 (\rho - i \eta) + \mc{O}(\lambda^4)\,.
\end{align} 
The CKM hierarchy suggests that the CP-averaged rate is linearly sensitive to $\re \ycc$. Indeed, we find that the rate is only quadratically sensitive to $\im \ycc$ and that $\yuu$ does not contribute up to $\mc{O}(\lambda^4)$, thus
\begin{align}
    \frac{d(\Gamma_K+\bar\Gamma_K)/2}{dq^2} &=
    \label{eq:BKll-AvgRate}
     \mc{N}_K \left(f^{(K)}_+\right)^2 \Bigg[\Cc_{10}^2 + \left(\Cc_9+\tilde{f}^{(K)}_T \Cc_7 \right)^2 \nonumber\\
     +2\bigg(\Cc_9 &+\tilde{f}^{(K)}_T \Cc_7 \bigg) \re\ycc + \mc{O}(|\ycc|^2,\lambda^4)\Bigg]\,. 
\end{align}
 Although we are considering low $q^2 \lesssim 6\e{GeV^2}$ region below the $c\bar{c}$ threshold, we cannot discard $\im \ycc$ since the $\bar c c\,\bar s b$ operators can 
 generate absorbtive contributions at any $q^2$ via the intermediate on-shell $D D_s^*$ and similar states~\cite{Khodjamirian:2012rm,Asatrian:2019kbk, Ciuchini:2021smi}. On the other hand, since the currently observed deviations of experimental rates from  predictions based on local SM contributions are up to $\sim 20\%$, we have linearized the dependence of the rate on $\ycc$. This makes our analysis valid in the regime $|\ycc| \ll |\Cc_9|$, which is supported by current experimental results barring accidental cancellations between large hypothetical new physics~(NP) and long distance contributions.
 
The CP-odd rate difference, being proportional to $\im  \tilde \lambda^{(s)}_{c} $, is $\lambda^2 \eta$ suppressed compared to the CP-averaged rate. At linear order in $\yuu ,\ycc$ the expression \eqref{eq:rateDiff-general} simplifies to
\begin{equation}
\label{eq:BKRateDif}
\begin{split}
  \frac{d(\Gamma_K-\bar\Gamma_K)}{dq^2} &  = 4 \mathcal{N}_K \left(f^{(K)}_+\right)^2 \,(\eta \lambda^2)\,\Big[ \im  (\yuu -\ycc)\\
    &\phantom{=} \times
    \left(\Cc_9+\tilde f^{(K)}_T \Cc_7 \right) + \mc{O}(\yuu \ycc) \Big]\,.
    \end{split}
\end{equation}

It is interesting to note that in this approximation the CP-odd rate difference in $B \to K \ell \ell$ is sensitive only to $\im(\yuu-\ycc)$. Consequently, in this region and assuming only SM short distance contributions, combined measurements of CP-averaged rates and CP-odd rate differences can completely determine the contributing long distance amplitudes, namely $\im  (\yuu - \ycc)$ and $\re  \ycc$. More generally however, presence of two independent long-distance quantities in two observables implies that BSM effects in $\Cc_7$ and/or $\Cc_9$ cannot be disentangled from effects in $\im  (\yuu - \ycc)$ and $\re  \ycc$ without additional theoretical input for the latter (for a recent attempt in this direction see Refs.~\cite{Gubernari:2022hxn, LHCb:2023gel, LHCb:2023gpo}).

\begin{figure}[!h]
    \centering
    \begin{tabular}{c}
    \includegraphics[width=0.95\columnwidth]{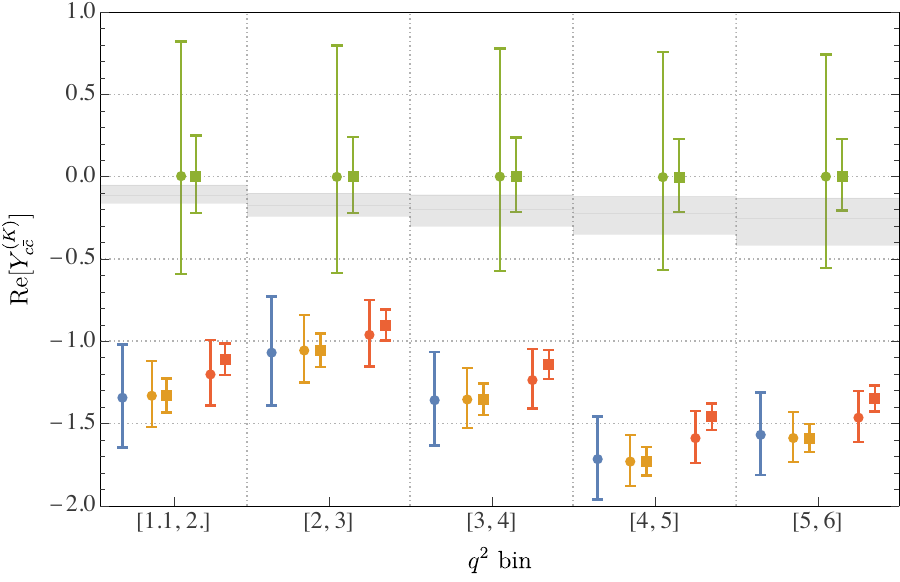} \\
    \includegraphics[width=0.95\columnwidth]{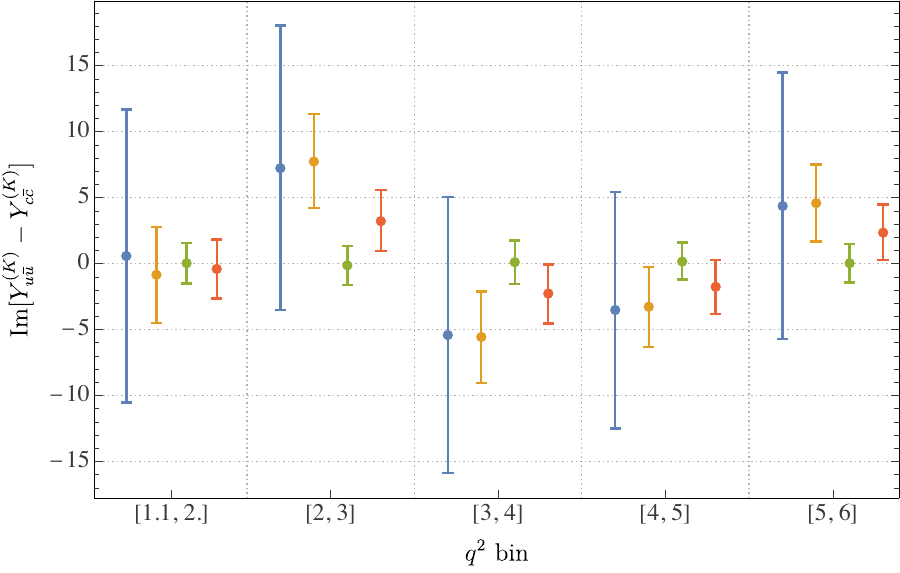} 
    \end{tabular}
   \caption{Current (LHCb) and projected $B \to K \mu \mu$ and $B\to\pi\mu\mu$ $1\sigma$ constraints on ${\rm Re}[\ycc^{(K)}]$ (top) and ${\rm Im}[\yuu^{(K)} -\ycc^{(K)}]$ (bottom) by $q^2$ bins from 1.1 to 6 $\mathrm{GeV}^2$ assuming only SM short distance contributions. In blue we show the current$B \to K \mu \mu$ LHCb constrains~\cite{LHCb:2014mit}. In yellow, projections for a factor 3 reduction in experimental uncertainties. In green, projected constrains from the $B\to\pi\mu\mu$ mode assuming a $20\%$ uncertainty for both the CP-averaged rate $\frac{d(\Gamma+\bar\Gamma)/2}{dq^2}$ and the direct CP-asymmetry ($\mathcal{A}_{\rm CP}$) measurements, and marginalizing over the $U$-spin breaking parameters $\epsilon_c$ and $\epsilon_{uc}$ defined in Sec.~\ref{sec:combine} assuming a uniform prior with support $[-0.3,0.3]$. In orange, we show the combination of both $B \to K \mu \mu$ and  $B\to\pi\mu\mu$ projected constrains.
   Constrains assuming current theoretical uncertainties are represented by circles, while a factor 3 reduction on them is represented by squares. Reduced theoretical uncertainties are not shown in the ${\rm Im}[\yuu^{(K)} -\ycc^{(K)}]$ plot since it is mostly dominated by experimental uncertainties. For comparison the predictions for ${\rm Re}[Y_{c\bar{c}}^{(K)}]$ for $B\to K\mu\mu$ from Ref.~\cite{Khodjamirian:2012rm} are shown as a gray band.}\label{fig:ReYccImYuumYccbyBins}
\end{figure}

To illustrate the sensitivity of current LHCb measurements, namely the binned CP-averaged rate and the direct CP asymmetry (defined as $\mathcal{A}_{CP} =  (\Gamma_K-\bar\Gamma_K)/(\Gamma_K + \bar\Gamma_K)$) in $B\to K \mu^+ \mu^-$~\cite{LHCb:2012kz, LHCb:2013lvw, LHCb:2014mit}, we plot in Fig.~\ref{fig:ReYccImYuumYccbyBins} in blue and yellow the resulting allowed parameter space in the plane of $\im (\yuu-\ycc)$ and $\re  \ycc$\footnote{More precisely we plot their values averaged over the corresponding $q^2$ bins.}, for the $q^2$ bins ranging from $q^2 \in [1.1,2]$~GeV${}^2$ to $q^2 \in [5,6]$~GeV${}^2$, and when assuming only SM short distance contributions. To derive these constraints we have used the most recent evaluation of the relevant form factors including their (correlated) uncertainties from Ref.~\cite{Gubernari:2023puw}, while the CKM and EW parameters as well as quark and meson masses are taken from PDG~\cite{ParticleDataGroup:2022pth}. 

 The currently allowed ranges from experimental data can be compared to theoretical estimates for $\ycc$ and $\yuu$ based on QCD factorisation and light cone sum rules (dubbed light cone operator product expansion -- LCOPE)~\cite{Khodjamirian:2012rm,Khodjamirian:2017fxg}, shown in gray.

In the case of $b\to d$ transition there is no hierarchy between the CKM factors
\begin{eqnarray}
    \tilde \lambda_c^{(d)} &=& \frac{\rho-1 + i \eta}{(1-\rho)^2 + \eta^2 } + \mc{O}(\lambda^2)  \,,\\
    \tilde \lambda_u^{(d)} &=& -1 - \frac{\rho-1+i\eta}{(1-\rho)^2+\eta^2} +\mc{O}(\lambda^2) \,,
\end{eqnarray}
however, the numerical values $\tilde\lambda_c^{(d)} = -0.98 +0.40\,i $, $\tilde\lambda_u^{(d)} = -0.020 -0.40\,i$ reveal an accidental cancellation in the real part of $\tilde\lambda_u^{(d)}$, which is related to the fact that unitarity angle $\alpha = \phi_3 \approx \pi/2$. Nonetheless, the CP-averaged rate depends in general on both $\ycc$, $\yuu$:
\begin{widetext}
\begin{align}
 \frac{d(\Gamma_\pi+\bar\Gamma_\pi)/2}{dq^2}
    &= 
    \mc{N}_\pi \left(f^{(\pi)}_+\right)^2 \left[\Cc_{10}^2 + (\Cc_9+\tilde{f}^{(\pi)}_T \Cc_7)^2 + 2 (\Cc_9 +\tilde{f}^{(\pi)}_T \Cc_7)\re \ycc +|\ycc|^2 +  (\im \tilde\lambda_u^{(d)})^2 |\yuu-\ycc|^2 + \cdots \right]\nonumber\\
     &\approx 
    \mc{N}_\pi \left(f^{(\pi)}_+\right)^2 \left[\Cc_{10}^2 + \left(\Cc_9+\tilde{f}^{(\pi)}_T \Cc_7\right)^2 + 2 \left(\Cc_9+\tilde{f}^{(\pi)}_T \Cc_7\right) \re \ycc  \right]\label{eq:dGammaPi} \,.
    \end{align} 
\end{widetext}
The omitted $\cdots$ terms in the first line are suppressed by $\rho(1-\rho)-\eta^2 = 0.022$. The leading dependence is again on $\ycc$ while $\yuu$ enters with a prefactor $(\im\tilde\lambda_u^{(d)})^2 \approx 0.16$. In the final result we have thus neglected terms with $(\im\tilde\lambda_u^{(d)})^2$ and linearized the expression in $\ycc$. Note that for theoretically preferred values of $\ycc$ and $\yuu$~\cite{Hambrock:2015wka}, the error we are making with the neglected terms in the last line of eq.~\eqref{eq:dGammaPi} is at most $\sim 2\%$\footnote{This error corresponds to $|\ycc| \sim 1$.}. The $B\to \pi$ CP-odd rate difference reads
\begin{align}
\label{eq:BpiRateDif}
  \frac{d(\Gamma_\pi-\bar\Gamma_\pi)}{dq^2} &  \approx 4 \mathcal{N}_\pi \left(f^{(\pi)}_+\right)^2 \,\frac{\eta}{(1-\rho)^2+\eta^2} \, 
  \nonumber\\
  & \phantom{\approx} \times \im  (\yuu - \ycc) \left(\Cc_9+\tilde f^{(\pi)}_T \Cc_7 \right)\,,
\end{align}
where we have truncated quadratic terms as in eq.~\eqref{eq:BKRateDif}.
{
Since currently there is no available determination of the $B\to \pi\mu\mu$ CP-asymmetry in the non-resonant $q^2$ regions, we show in Fig.~\ref{fig:ReYccImYuumYccbyBins} projected constrains of ${\rm Re}[\ycc]$  and ${\rm Im}[\yuu -\ycc]$ assuming a $20\%$ uncertainty for both the CP-averaged rate and the direct CP-asymmetry. In order to combine and compare the sensitivity of extractions from $B\to K$ and $B\to \pi$ modes, we introduce two U-spin breaking parameters $\epsilon_c$ and $\epsilon_{uc}$ (see Sec.~\ref{sec:combine}) and treat them as nuisance parameters sampled uniformly in the $[-0.3,0.3]$ interval. 
}

%
\section{Combining $B\to K$ and $B\to \pi$ CP-differences}
\label{sec:combine}
%

Given the similarity between the theoretical expressions for the CP-odd decay rate differences in $B\to K$ (eq.~\eqref{eq:BKRateDif}) and $B\to \pi$  (eq.~\eqref{eq:BpiRateDif}) transitions, it is interesting to consider their ratio 
\begin{equation}
\label{eq:CPratio}
R^\mrm{CP}_{K/\pi} \equiv \frac{(d\Gamma_K-d\bar\Gamma_K)/dq^2}{(d\Gamma_\pi-d\bar\Gamma_\pi)/dq^2}\,.
\end{equation}
In the $U$-spin limit of $Y_{q\bar q}$ functions, namely that $\yuu(q^2)$ and $\ycc(q^2)$ are equal for the $B \to K$ and $B\to \pi$ decays\footnote{Namely, that $\yuu^{(K)}(q^2) = \yuu^{(\pi)}(q^2)$ and $\ycc^{(K)}(q^2) = \ycc^{(\pi)}(q^2)$.}, the $R^\mrm{CP}_{K/\pi}$ ratio is predictable within the SM. We do incorporate well-known sources of $U$-spin breaking in the form factors and kinematics. The ratio is $\mc{O}(1)$, since both rate differences are of Cabibbo order $A^2\lambda^6$, even though CP-averaged rates and CP-asymmetries are very different for the two modes.
This time we use the expression~\eqref{eq:rateDiff-general} including quadratic terms in $\ycc,\yuu$ and obtain the SM prediction
for the ratio
\begin{align}
\label{eq:CPratioSM}
R^\mrm{CP}_{K/\pi}\big|_\mrm{SM} 
  &=\left(\frac{\lambda_K}{\lambda_\pi}\right)^{3/2}
  \left(\frac{f^{(K)}_+}{f^{(\pi)}_+}\right)^2
   \nonumber\\
   & \phantom{=} \times \left[1+\frac{\Cc^\mrm{SM}_7 \left(\tilde f^{(K)}_T-\tilde f^{(\pi)}_T \right)}{\Cc^\mrm{SM}_9 + \Cc^\mrm{SM}_7 \tilde f^{(\pi)}_T} \right] (1-\Delta_U)\,. 
\end{align}
The net total $U$-spin breaking corrections and quadratic $\yqq$ terms are absorbed in
\begin{equation}
    \Delta_U = \epsilon_{uc} + \frac{\im\left(\ycc^{(K)} \yuu^{(K)*}\right)}{\im(\yuu^{(K)}-\ycc^{(K)})}\,    
    \frac{\Cc_7^\mrm{SM} (\tilde f^{(K)}_T-\tilde f^{(\pi)}_T)}{\left(\Cc^\mrm{SM}_9+\Cc^\mrm{SM}_7 \tilde f^{(K)}_T\right)^2}\,.
\end{equation}
 Here we have used $\yuu^{(K)}$ and $\ycc^{(K)}$ as four-quark contributions in $B \to K \ell \ell$ whereas the corresponding $B \to \pi \ell \ell$ contributions are given as $\im(\yuu^{(\pi)}-\ycc^{(\pi)}) = \im(\yuu^{(K)}-\ycc^{(K)}) (1+\epsilon_{uc})$.
The $U$-spin breaking $\epsilon_{uc}$ can be up to $\sim 30\%$, a value supported by the experimental data on branching fractions of $B^+ \to J/\psi K^+$ and $B^+ \to J/\psi \pi^+$. Namely, the amplitudes of those decays are proportional to $\ycc^{(K)}(q^2=m_{J/\psi}^2)$ and $\ycc^{(\pi)}(q^2=m_{J/\psi}^2)$, respectively. We extract their ratio from the measured decay widths~\cite{ParticleDataGroup:2022pth}, and while correcting for differences in CKM factors and final state momenta $|\bm{k}_P|$, we find that the $U$-spin breaking is indeed within the assumed limits:
\begin{align}
    \left|\frac{\ycc^{(K)}}{\ycc^{(\pi)}}\right|_{q^2=m_{J/\psi}^2} &= \left|\frac{\lambda^{(d)}_c}{\lambda^{(s)}_c}\right|\sqrt{\frac{|\bm{k}_\pi|}{|\bm{k}_K|}\frac{\Gamma(B^+ \to J/\psi K^+)}{\Gamma(B^+ \to J/\psi \pi^+)}} \nonumber\\
    &\simeq 1.2 \,.
\end{align}
 The second term in $\Delta_U$ is numerically suppressed by $\Cc_7/\Cc_9^2$ and $\tilde f_{T}^{(K)} - f_{T}^{(\pi)}$ and is subleading with respect to $\epsilon_{uc}$.\footnote{Note also that $\im \ycc$ and $\im \yuu$ are expected to be significantly different functions of $q^2$ and thus the prefactor of the second term is expected to be well bounded.} Thus the uncertainty on $R^\mrm{CP}_{K/\pi}\big|_\mrm{SM}$ is determined by $U$-spin breaking while the values of $\yqq$ play no important role.

\begin{figure}[!h]
    \centering\includegraphics[width=0.95\columnwidth]{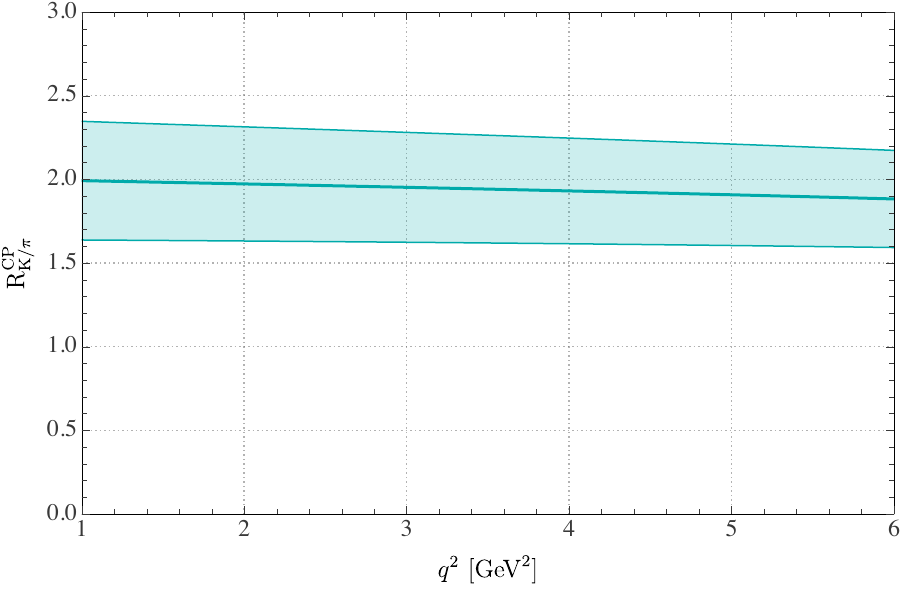} 
   \caption{Standard Model prediction for the $R^\mrm{CP}_{K/\pi}$ ratio in the U-spin limit for the hadronic corrections ($\Delta_U=0$).} 
   \label{fig:RCPKpiSM_USpinLimit}
\end{figure}

\begin{table}[]
\def\arraystretch{1.5}
\begin{tabular}{c|c}
                $q^2 [{\rm GeV}]$ bin& $R^\mrm{CP}_{K/\pi}$   \\\hline 
                $[1.1,2.0]$&  $1.92\pm 0.32$  \\
                $[2.0,3.0]$&  $1.90\pm 0.31$ \\
                $[3.0,4.0]$&  $1.88\pm 0.30$ \\
                $[4.0,5.0]$&  $1.86\pm 0.29$  \\
                $[5.0,6.0]$&  $1.84\pm 0.28$ \\ 
\end{tabular}
\caption{Binned Standard Model prediction for the $R^\mrm{CP}_{K/\pi}$ ratio in the U-spin limit for the hadronic corrections ($\Delta_U=0$).}
\end{table}
The square bracket in eq.~\eqref{eq:CPratioSM} should be close to 1 due to 
smallness of both $\Cc_7^\mrm{SM}$ and known $U$-spin breaking $\tilde f^{(K)}_T-\tilde f^{(\pi)}_T$. Note also that the dependence on CKM cancels out in eq.~\eqref{eq:CPratioSM} and this expression remains valid in the presence of MFV new physics where the contributions to $b\to s \ell \ell$ and $b \to d \ell \ell$ are aligned with the CKM factors of the SM~\cite{Bause:2022rrs}. Specifically, the contribution of new physics  to $\Cc_9$ and $\Cc_7$ should be real and $U$-spin-symmetric in order for \eqref{eq:CPratioSM} to apply, with appropriate real shifts $\Cc_i^\mrm{SM} \to \Cc_i^\mrm{SM} + \delta \Cc_i$. 

In order to incorporate more general scenarios, such as having NP contribution only in $b\to s$ transitions, we have to modify the $R^\mrm{CP}_{K/\pi}$ expression accordingly. Here we introduce NP as a real modification of Wilson coefficients for $b\to s\ell \ell$: $\Cc^{(s)}_{7,9} = \Cc^\mrm{SM}_{7,9} + \delta \Cc_{7,9}^{(s)}$. For such $U$-spin breaking NP contributions $R^\mrm{CP}_{K/\pi}$ becomes
\begin{align}
\label{eq:RKpiNP}
R^\mrm{CP}_{K/\pi}\big|_\mrm{ReNP}
  &= R^\mrm{CP}_{K/\pi}\big|_\mrm{SM} \left(1+\frac{\delta \Cc_9^{(s)}+\delta \Cc_7^{(s)} \tilde f_T^{(K)}}{\Cc^\mrm{SM}_9 + \Cc^\mrm{SM}_7 \tilde f^{(K)}_T}\right)\,.
\end{align}
In order to discern $\delta \Cc_9^{(s)}$ effect from the $U$-spin breaking the relative modification of $\delta \Cc_9^{(s)}/\Cc_9^\mrm{SM}$ should be larger than $\epsilon_{uc}$ and form factors' uncertainty combined.

\subsection{CP-violating new physics}
Let us discuss now what happens if we instead introduce CPV NP contributions to Wilson coefficients, $\delta \Cc_i = i\, \im \,\delta \Cc_i$. The CP-odd differences of $B \to K$ rates and $B\to \pi$ get modified by additional contributions, on top of those in~\eqref{eq:BKRateDif} and~\eqref{eq:BpiRateDif}:
\begin{align}
\label{eq:BKRateDifNP}
  \delta\left(\frac{d(\Gamma_K-\bar\Gamma_K)}{dq^2}\right)_\mrm{ImNP} &\approx 4 \mathcal{N}_K \left(f^{(K)}_+\right)^2 \,\im\ycc^{(K)} \nonumber\\
  & \quad \times \left(\im \delta\Cc_9^{(s)} + \tilde f^{(K)}_T \im \delta\Cc^{(s)}_7 \right)\,,\\
  \delta\left(\frac{d(\Gamma_\pi-\bar\Gamma_\pi)}{dq^2}\right)_\mrm{ImNP} &\approx 4 \mathcal{N}_\pi \left(f^{(\pi)}_+\right)^2 \,\im\ycc^{(\pi)}\,\nonumber\\
  & \quad \times \left(\im \delta\Cc_9^{(d)} + \tilde f^{(\pi)}_T \im \delta\Cc^{(d)}_7 \right)\,.
\end{align}
These contributions are proportional to CP-even real parts $\re\tilde \lambda_{u,c}^{(q')}$ whose Cabibbo hierarchy selects $\im \ycc^{(P)}$ 
as a dominant source of the strong phase~\cite{Becirevic:2020ssj}. Consequently CPV NP in $R_{K/\pi}^\mrm{CP}$ is affected by additional $U$-spin breaking through $\im \ycc^{(\pi)} = (1+\epsilon_c) \im \ycc^{(K)}$. The relative modification of $R_{K/\pi}^\mrm{CP}\big|_\mrm{SM}$ is
\begin{widetext}
    \begin{equation}
       \frac{\delta R_{K/\pi}^\mrm{CP}\big|_\mrm{ImNP}}{R_{K/\pi}^\mrm{CP}\big|_\mrm{SM}} \approx 
  \frac{\im \ycc^{(K)}}{\im(\yuu^{(K)}-\ycc^{(K)})}\left(\frac{\frac{1}{\eta \lambda^2} (\im \delta \Cc_9^{(s)} +  \tilde f^{(K)}_T \im \delta \Cc_7^{(s)})}{\Cc_9^\mrm{SM}+ \tilde f^{(K)}_T \Cc^\mrm{SM}_7} - \frac{\frac{(1-\rho)^2+\eta^2}{\eta} (\im \delta \Cc_9^{(d)} +  \tilde f^{(\pi)}_T \im \delta \Cc_7^{(d)})}{\Cc_9^\mrm{SM}+ \tilde f^{(\pi)}_T \Cc^\mrm{SM}_7}\right) \,,
    \end{equation}
\end{widetext}
where we have neglected the $U$-spin breaking terms $\epsilon_c$ and $\epsilon_{cu}$ and only show the leading effect in presence of purely imaginary NP Wilson coefficients. 
The weight in front of the $\im \delta \Cc_i^{(s)}$ is large -- $1/(\eta \lambda^2) = 54$ -- due to relative smallness of CP-violation in $b \to s \ell \ell$ in the SM.

%
\section{Discussion and Conclusions}
\label{sec:conclusions}

Effects of four-quark operators in rare semileptonic $B$ meson decays have important implications both for our understanding of QCD dynamics as well as for physics BSM. In this work we have shown how combined measurements of CP-averaged decay rates and direct CP-asymmetries of $B^\pm\to K^\pm \ell^+ \ell^-$ and $B^\pm\to \pi^\pm \ell^+ \ell^-$ transitions can be used to learn about the sizes of the four-quark operator matrix elements and how their contributions interplay with possible local BSM effects. 

The relative importance of dispersive and absorptive amplitudes involving $ \bar b q'\bar c c$ and $\bar b q'  \bar u u$ operators in the SM, dictated by the hierarchies of the CKM, leads to an interesting interplay of CP-even and CP-odd observables in $B\to K$ and $B\to \pi$ transitions. In particular, our results motivate dedicated binned measurements of the direct CP asymmetry (or the corresponding CP-odd decay rate difference) in $B\to \pi \ell^+ \ell^-$. As shown in Fig.~\ref{fig:ReYccImYuumYccbyBins} such measurements could help significantly reduce the current uncertainty on the absorptive long distance amplitudes entering $B\to K \ell^+ \ell^-$. The corresponding improvement in precision of dispersive amplitudes is projected to be modest, with one caveat: in our projections we have assumed that the experimental precision on $B\to K$ modes will remain an order of magnitude better compared to measurements with pion final states. 
Within our approach, current measurements are in mild tension
with existing theoretical estimates~\cite{Khodjamirian:2012rm} of the dominant dispersive long distance SM contributions to $B\to K \ell^+ \ell^-$ rates. A similar comparison for absorptive amplitudes would require a dedicated theoretical estimation of $ \bar b q'  \bar u u$ operator effects, which is currently not available in the literature and beyond the scope of this work. We also note that the extraction of absorptive LD amplitudes from the CP-odd rate difference measurements will remain dominated by experimental uncertainties even for our HL-LHCb projections and thus insensitive to theoretical form factor uncertainties. 

Motivated by these results, we have also constructed a ratio $R^\mrm{CP}_{K/\pi} $ of CP-odd decay rate differences of $B^\pm\to K^\pm \ell^+ \ell^-$ and $B^\pm\to \pi^\pm \ell^+ \ell^-$ which can be computed exactly in the $U$-spin limit of the SM, see eq.~\eqref{eq:CPratioSM}. We have estimated $R^\mrm{CP}_{K/\pi} $ in presence of known $U$-spin breaking effects due to kinematics and differences in form factors. On the other hand, the explicit dependence on LD amplitudes is suppressed by $U$-spin breaking and $|\Cc_7 / \Cc_9^2| \lesssim 0.02$. The current theoretical uncertainties on $R^\mrm{CP}_{K/\pi} |_{\rm SM}$ are currently dominated by our knowledge of the relevant form factors, mainly because we have to treat their uncertainties as completely uncorrelated. Conversely, a correlated extraction of $B\to K$  and $B\to \pi$  form factors on the Lattice could potentially significantly reduce this error. Furthermore, the remaining $U$-spin breaking effects in LD amplitudes could in principle be estimated, for example using light-cone sum rules techniques~\cite{Khodjamirian:2012rm, Hambrock:2015wka}. Thus, $R^\mrm{CP}_{K/\pi} $ has the potential to become one of the theoretically cleanest observables related to rare semileptonic $B$ meson decays within the SM.

We have also explored the interplay between SM four-quark contributions and possible short-distance NP effects in CP-odd $B\to P \ell^+ \ell^-$ decay rate differences. In presence of CP conserving NP affecting only the kaon mode, one could in principle disentangle it from SM LD effects using the pion mode measurements. In particular, the extraction of absorptive amplitudes from both modes in this case would disagree (i.e. green and yellow bars in Fig.~\ref{fig:ReYccImYuumYccbyBins}). Obviously, such a discrimination is in practice observable only if the relevant NP effects are bigger than uncertainties related to $U$-spin breaking. In case of CPV NP, larger relative effects are expected in the $B\to K$ mode, due to the CKM suppression of CPV within the SM. Unfortunately, quantitative predictions of such NP effects on $R^\mrm{CP}_{K/\pi} $ would require independent theoretical estimations of the absorptive $\bar b q'\bar c c$ four-quark amplitudes (i.e. they cannot be extracted from measurements of the $B\to \pi$ transition). 

We conclude by noting that, while this work focused on $B\to P \ell^+ \ell^-$ transitions, the same analysis can be applied to individual helicity amplitudes in rare semileptonic decays of B-mesons to vector meson final states (e.g. $B\to \rho {\rm ~vs.~} B\to K^*$ or $B_s \to \phi {\rm ~vs.~} B_s \to K^*$). We leave a dedicated analysis of such transitions for future work.

\begin{acknowledgments}
JFK and NK acknowledge financial support from the Slovenian Research Agency (research core funding No. P1-0035 and J1-3013). MN acknowledges the financial support by the Spanish Government (Agencia Estatal de Investigaci\'on MCIN/AEI/10.13039/501100011033)  and the European Union NextGenerationEU/PRTR through the “Juan de la Cierva” program (Grant No. JDC2022-048787-I)
and Grant No. PID2020-114473GB-I00. MN also acknowledges the support of the Generalitat Valenciana through Grant No. PROMETEO/2021/071. This study has been partially carried out within the INFN project (Iniziativa Specifica) QFT-HEP. NK acknowledges support of the visitor programme at CERN Department of Theoretical Physics, where part of this work was done.
\end{acknowledgments}
\bibliography{main}

\end{document}